\documentstyle[aps,prb,preprint,tighten]{revtex}

\begin{document}

\title{Electronic polarization in the
ultrasoft pseudopotential formalism}

\author{David Vanderbilt}
\address{Department of Physics and Astronomy, Rutgers University, 
Piscataway, NJ 08855-0849, USA}
\author{R.D.~King-Smith}
\address{Molecular Simulations Inc., 9685 Scranton Road,
San Diego, CA 92121, USA}

\date{January 17, 1998}

\maketitle
\begin{abstract}

An expression is given for the electronic polarization of an
insulating crystal within the ultrasoft pseudopotential scheme.
The pseudopotential charge-augmentation terms modify the usual
Berry-phase expression, and also give rise to a second term that
takes the form of a conventional expectation value.

\end{abstract}


\mediumtext


Several years ago, we presented a formulation that
allows for the calculation of the electronic polarization of an
insulating crystal in the context of a
conventional electronic band-structure calculation.\cite{ksv,vks}
It was shown that the contribution of a given band to the electronic
polarization could be expressed in terms of the charge center of its
associated Wannier function; or equivalently, in terms of the Berry
phase of the cell-periodic Bloch function $u_{\bf k}$ as the
wavevector $\bf k$ is adiabatically transported by a reciprocal
lattice vector.  Resta has given a review of this
theory.\cite{resta}

The above theory can be applied straightforwardly in the context of
an all-electron (e.g, augmented plane-wave) scheme, or any
norm-conserving pseudopotential scheme.  However, in the ultrasoft
pseudopotential scheme,\cite{usdv,uslaas,usbig} the
pseudo-wavefunctions do not obey a conventional normalization
condition, and the Berry-phase expression must be modified.

The purpose of this Report is to present the appropriate
formulation of the electronic polarization within the ultrasoft
pseudopotential scheme.  The expression has already been used in
applications,\cite{zhong,bernar} but has never previously been
reported.

Let $l=\{{\bf R},\tau\}$ label the atomic sites in the crystal,
where $\bf R$ is a real-space lattice vector and $\tau$ is the
extra fractional translation needed to specify the location of
the atom in the cell.  Note that $\tau$ will double as a species
label for the atom.  The basic ingredients of the USPP scheme
\cite{usdv,uslaas,usbig} are the projectors
$\beta_{\tau,i}({\bf r})$
and the charge augmentation functions
$Q_{\tau,ij}({\bf r})$ for an atom of species $\tau$.
Here $i$ is a composite index labeling the angular momentum $(lm)$
and radial indices of the projectors.  The matrix quantities
$Q_{\tau,ij}$ and ${\bf d}_{\,\tau,ij}$ are then defined via
\begin{equation}
Q_{\tau,ij}=\int d{\bf r}\,Q_{\tau,ij}({\bf r})
\label{eq:Qdef}
\end{equation}
and
\begin{equation}
{\bf d}_{\,\tau,ij}=\int d{\bf r}\,{\bf r}\,Q_{\tau,ij}({\bf r}) \;\;.
\label{eq:ddef}
\end{equation}
The projectors and charge augmentation functions of atom $l$ located
in the crystal at $\bf R+\tau$ are
\begin{equation}
\beta_i^{(l)}({\bf r})=\beta_{\tau,i}({\bf r-R-\tau})
\label{eq:bxtal}
\end{equation}
and
\begin{equation}
Q_{ij}^{(l)}({\bf r})=Q_{\tau,ij}({\bf r-R-\tau}) \;\;.
\label{eq:Qxtal}
\end{equation}
The charge density operator in the crystal is thus\cite{uslaas}
\begin{equation}
K({\bf r})=  {\bf\vert r\rangle\langle r\vert} +
\sum_l \sum_{ij} \> Q_{ij}^{(l)}({\bf r})\>
\vert\beta_i^{(l)}\rangle\,\langle\beta_j^{(l)}\vert
\label{eq:Kdef}
\end{equation}
and the number operator is $S=\int d{\bf r}\,K(\bf r)$ or
\begin{equation}
S=1+
\sum_l \sum_{ij} Q_{\tau,ij}\,
\vert\beta_i^{(l)}\rangle\,\langle\beta_j^{(l)}\vert \;\;.
\label{eq:Sdef}
\end{equation}

The Bloch pseudo-wavefunctions $\psi_{\bf k}$ are written
in terms of cell-periodic functions $u_{\bf k}$ in the usual
way,
\begin{equation}
\psi_{\bf k}({\bf r}) = e^{i\bf k\cdot r} \, u_{\bf k}({\bf r}) \;\;.
\end{equation}
The normalization
$\langle\psi_{\bf k}\vert S\vert\psi_{\bf k}\rangle_{_V} =1$
is relative to the number operator of Eq.~(\ref{eq:Sdef}).
(The notation $\langle\rangle_{_V}$ indicates that the expectation
value or matrix element is taken per unit cell of volume $V$.)
Substituting Eq.~(\ref{eq:Sdef}) and using the definition
\begin{equation}
\beta_{\tau,i}^{(\bf k)}({\bf r}) = 
    e^{-i\bf k\cdot(r-\tau)} \> \beta_{\tau,i}({\bf r-\tau}) \;\;,
\label{eq:betak}
\end{equation}
the normalization of the cell-periodic Bloch function becomes
\begin{equation}
1 = \langle u_{\bf k}\vert u_{\bf k}\rangle_{_V} +
\sum_\tau \sum_{ij} \, Q_{\tau,ij} \>
   \langle u_{\bf k} \vert \beta_{\tau,i}^{(\bf k)} \rangle \,
   \langle \beta_{\tau,j}^{(\bf k)} \vert u_{\bf k} \rangle
\;\;.
\label{eq:unorm}
\end{equation}

Now, from the Bloch pseudo-wavefunctions, one can construct a
localized Wannier pseudo-wavefunction in the usual way,
\begin{equation}
\vert w\rangle = {V\over(2\pi)^3} \int_{\rm BZ} d{\bf k}\,
\vert \psi_{\bf k}\rangle \;\;,
\label{eq:wdef}
\end{equation}
where the integral is over the Brillouin zone (BZ).  It is
straightforward to check the normalization
$\langle w\vert S\vert w\rangle=1$, viz.,
\begin{eqnarray}
\langle w\vert S\vert w\rangle &=&
\langle w\vert w\rangle + \sum_l \sum_{ij} \, Q_{\tau,ij} \>
   \langle w\vert\beta_i^{(l)}\rangle \,
   \langle\beta_j^{(l)}\vert w\rangle   \nonumber \\
&=& {V^2\over(2\pi)^6}
\int_{\rm BZ} d{\bf k} \int_{\rm BZ} d{\bf k'} \>
\sum_{\bf R} \>
\Biggl\{ \>
   \int_V d{\bf r}\> u_{\bf k}^*({\bf r}) \, u_{\bf k'}({\bf r}) \,
      e^{-i\bf(k-k')\cdot(R+r)}
 \nonumber \\
   && \qquad\qquad\qquad +\sum_\tau \sum_{ij} \, Q_{\tau,ij}  \>
      \langle u_{\bf k} \vert \beta_{\tau,i}^{(\bf k)} \rangle \,
      \langle \beta_{\tau,j}^{(\bf k')} \vert u_{\bf k'} \rangle
      \> e^{-i\bf(k-k')\cdot(R+\tau)}
\Biggr\}
\nonumber \\
&=& {V\over(2\pi)^3} \int_{\rm BZ} d{\bf k} \>
\Biggl\{ \>
\langle u_{\bf k}\vert u_{\bf k}\rangle_{_V} +
\sum_\tau \sum_{ij} \, Q_{\tau,ij} \>
   \langle u_{\bf k} \vert \beta_{\tau,i}^{(\bf k)} \rangle \,
   \langle \beta_{\tau,j}^{(\bf k)} \vert u_{\bf k} \rangle
\Biggr\} \;\;.
\label{eq:wnorm}
\end{eqnarray}
The second line results from substituting Eqs.~(\ref{eq:Sdef})
and (\ref{eq:wdef}), and the third is obtained from the completeness
relation
\begin{equation}
\sum_{\bf R} e^{-i\bf(k-k')\cdot R}={(2\pi)^3\over V} \> \delta\bf(k-k') \;\;.
\label{eq:complete}
\end{equation}
Using the Bloch normalization of Eq.~(\ref{eq:unorm}), it then follows
that $\langle w\vert S\vert w\rangle=1$.

The contribution of this band to the electronic polarization
is given in terms of the Wannier center as $(-2e/V)\langle\bf r\rangle$,
assuming spin-paired electrons.\cite{ksv,vks,resta}  Thus, our
remaining task is to calculate the location $\langle\bf r\rangle$
of the Wannier center.  Using Eqs.~(\ref{eq:Qdef}),
(\ref{eq:ddef}), (\ref{eq:Qxtal}), and (\ref{eq:Kdef}),
\begin{eqnarray}
\langle{\bf r}\rangle &=& \int d{\bf r}\,{\bf r}\,
   \langle w\vert K({\bf r}) \vert w\rangle \nonumber \\
&=& \langle w\vert{\bf r}\vert w\rangle
+\sum_l \sum_{ij} \>
   \Bigl[ {\bf(R+\tau)} Q_{\tau,ij} + {\bf d}_{\,\tau,ij} \Bigr] \>
   \langle w\vert\beta_i^{(l)}\rangle \,
   \langle\beta_j^{(l)}\vert w\rangle
   \;\;.
\label{eq:r1}
\end{eqnarray}
Substituting Eq.~(\ref{eq:wdef}), one obtains, in analogy with
the second line of Eq.~(\ref{eq:wnorm}),
\begin{eqnarray}
\langle{\bf r}\rangle &=&
{V^2\over(2\pi)^6}
\int_{\rm BZ} d{\bf k} \int_{\rm BZ} d{\bf k'}
\, \sum_{\bf R} \>
\Biggl\{
   \int_V d{\bf r}\, u_{\bf k}^*({\bf r}) \, {\bf(R+r)} \,  u_{\bf k'}({\bf r})
      \, e^{-i\bf(k-k')\cdot(R+r)}
 \nonumber \\
& & \quad
   +\sum_\tau \sum_{ij} \> [ {\bf(R+\tau)}Q_{\tau,ij} +{\bf d}_{\,\tau,ij}] \>
      \langle u_{\bf k} \vert \beta_{\tau,i}^{(\bf k)} \rangle \,
      \langle \beta_{\tau,j}^{(\bf k')} \vert u_{\bf k'} \rangle \,
      e^{-i\bf(k-k')\cdot(R+\tau)}
\Biggr\} \;\;.
\label{eq:r2}
\end{eqnarray}
Replacing $\bf R+r$ or $\bf R+\tau$ by $-i\nabla_{\bf k'}$ acting
on the exponential factor, then using an integration by parts to
cause $\nabla_{\bf k'}$ to act on the $u_{\bf k'}$ and
$\beta^{\bf(k')}$ instead, and applying the completeness relation
(\ref{eq:complete}), one finds
\begin{equation}
\langle{\bf r}\rangle =
\langle{\bf r}\rangle^{\rm BP} +
\langle{\bf r}\rangle^{\rm EV}
\label{eq:bpev}
\end{equation}
where
\begin{equation}
\langle{\bf r}\rangle^{\rm BP} =
   {V\over(2\pi)^3} \int_{\rm BZ} d{\bf k} \, {\bf A(k)} \;\;,
\label{eq:bp}
\end{equation}
\begin{equation}
\langle{\bf r}\rangle^{\rm EV} =
{V\over(2\pi)^3} \int_{\rm BZ} d{\bf k} \>
\sum_\tau \sum_{ij} \,{\bf d}_{\,\tau,ij} \,
   \langle u_{\bf k} \vert \beta_{\tau,i}^{(\bf k)} \rangle \,
   \langle \beta_{\tau,j}^{(\bf k)} \vert u_{\bf k} \rangle
   \;\;,
\label{eq:ev}
\end{equation}
and
\begin{equation}
{\bf A(k)} =
i\,\langle u_{\bf k}\vert \nabla_{\bf k}\vert u_{\bf k}\rangle_{_V}
+ \sum_\tau \sum_{ij} i\,Q_{\tau,ij} \,
   \langle u_{\bf k} \vert \beta_{\tau,i}^{(\bf k)} \rangle
   \,\nabla_{\bf k}\,
   \langle \beta_{\tau,j}^{(\bf k)} \vert u_{\bf k} \rangle
   \;\;.
\label{eq:Adef}
\end{equation}

The decomposition (\ref{eq:bpev}) is into ``Berry-phase'' (BP)
and ``expectation-value'' (EV) terms.  The latter involves
no coupling between k-points [cf.\ Eq.~(\ref{eq:ev})]
and is thus obviously gauge-invariant,
i.e., invariant with respect to a k-dependent change of
phase of the $u_{\bf k}$.  The BP term is also gauge-invariant,
as can be seen by substituting
$u_{\bf k}\rightarrow u_{\bf k} \times \exp[-i\gamma\bf(\bf k)]$
into (\ref{eq:Adef}) and using the Bloch normalization condition
(\ref{eq:unorm}) to show that
${\bf A(k)}\rightarrow{\bf A(k)}+\nabla_{\bf k}\gamma({\bf k})$.
The quantity $\bf A(k)$ thus plays the role of the ``Berry
connection'' or ``gauge potential'' of the Berry-phase
theory.\cite{berry,resta96}  Since $\exp[-i\gamma\bf(\bf k)]$ is
periodic in k-space, it follows that $\gamma({\bf k})$ must take
the form of $\bf R\cdot k$ plus a periodic part ($\bf R$ being a
lattice vector), so that $\langle{\bf r}\rangle^{\rm BP}$ is
clearly invariant modulo a lattice vector.

In practical calculations, a discrete mesh of k-points is
typically used.  The expectation-value term is easily evaluated
on any mesh of $N_k$ k-points as
\begin{equation}
\langle{\bf r}\rangle^{\rm EV} = {1\over N_k}
\sum_{\bf k} \sum_\tau \sum_{ij} \, {\bf d}_{\,\tau,ij} \,
   \langle u_{\bf k} \vert \beta_{\tau,i}^{(\bf k)} \rangle \,
   \langle \beta_{\tau,j}^{(\bf k)} \vert u_{\bf k} \rangle \;\;.
\label{eq:evdis}
\end{equation}
Following Ref.~\onlinecite{ksv}, the Berry-phase term is
evaluated one component at a time.  That is, the component
of $\langle{\bf r}\rangle^{\rm BP}$ along some primitive
reciprocal lattice vector $\bf G$ is evaluated by choosing
a 2D mesh of $N_{k\perp}$ strings of k-points running
parallel to $\bf G$, and calculating
\begin{equation}
\langle r_\parallel\rangle^{\rm BP} = {-1\over GN_{k\perp}}
\sum_{\bf k_\perp} \, {\rm Im}\,\ln\,\prod_{k_\parallel} \,
\Biggl\{
\langle u_{\bf k}\vert u_{\bf k+b}\rangle_{_V}
+\sum_\tau \sum_{ij} \, Q_{\tau,ij} \>
   \langle u_{\bf k} \vert \beta_{\tau,i}^{(\bf k)} \rangle \,
   \langle \beta_{\tau,j}^{(\bf k+b)} \vert u_{\bf k+b} \rangle
\Biggr\} \;\;.
\label{eq:bpdis}
\end{equation}
Here $r_\parallel$ and $k_\parallel$ are components of $\bf r$
and $\bf k$ parallel to $\bf G$, with ${\bf k}={\bf
k_\perp}+k_\parallel\hat{\bf G}$, and $\bf b$ being the separation
between neighboring k-points along the strings.  As in
Ref.~\onlinecite{ksv}, this discrete version of the Berry-phase
term is manifestly invariant with respect to a change of phase
of any one of the $u_{\bf k}$, since each $u_{\bf k}$ appears
once in a bra and once in a ket.

For the case of multiple discrete bands, the final expression
for the polarization is
\begin{equation}
{\bf P} = {\bf P}_{\rm ion} - {2e\over V} \sum_n \, \Bigl[
   \langle{\bf r}\rangle_n^{\rm BP}
 + \langle{\bf r}\rangle_n^{\rm EV}
\Bigr] \;\;,
\label{eq:pol}
\end{equation}
where the sum is over occupied bands $n$.  For the case of
compound multiple bands that have degeneracies at certain
locations within the Brillouin zone, the EV terms present
no difficulties, but the BP evaluation requires care.
One can again follow the example of Ref.~\onlinecite{ksv}
to express the Berry-phase contribution to the electronic
polarization as
\begin{equation}
{\bf P}^{\,\rm BP} = {-2e\over V}\,{1\over GN_{k\perp}}\,
\sum_{\bf k_\perp} \, {\rm Im}\,\ln\,\prod_{k_\parallel}
{\rm det}\,M^{\bf(k)}
\label{eq:bpmult}
\end{equation}
where
\begin{equation}
M^{\bf(k)}_{mn}=
\langle u_{m,\bf k}\vert u_{n,\bf k+b}\rangle_{_V}
+\sum_\tau \sum_{ij} \, Q_{\tau,ij} \>
   \langle u_{m\bf k} \vert \beta_{\tau,i}^{(\bf k)} \rangle \,
   \langle \beta_{\tau,j}^{(\bf k+b)} \vert u_{n,\bf k+b} \rangle
\label{eq:Mdef}
\end{equation}
and the indices $m$ and $n$ run over occupied bands.

In summary, the formulas for electronic polarization given in
Ref.~\onlinecite{ksv} have been generalized to the case in which
ultrasoft pseudopotentials have been used in the band-structure
calculation.  The resulting formulas are easily implemented
in practice.

\acknowledgments

This work was supported by NSF grant DMR-96-13648.
We thank V.~Fiorentini for encouraging the writing of
the manuscript.



\end{document}